\documentclass[12pt,twoside]{article}
\usepackage{fleqn,espcrc1}

\usepackage{graphicx}
\usepackage[figuresright]{rotating}

\title{The Role of Z=0 AGB Stars on the Early Chemical Enrichment} 

\author{I. Dom\'\i nguez\address[UGR]{Universidad de Granada, Granada, 
   Spain}, 
     C. Abia\addressmark[UGR], 
        O. Straniero\address{Osservatorio Astronomico di Collurania, 
      Teramo, Italy}, 
       A. Chieffi\address{Istituto di Astrofisica Spaziale (CNR), Roma, 
      Italy} 
        and 
  M. Limongi\address{Osservatorio Astronomico di Roma, 
                 Monteporzio, Italy}} 
       
\begin{document}

\maketitle

\begin{abstract}
 
We present theoretical evolutionary models for $Z=0$ stars in the mass
range $4\leq M/M_\odot\leq 8$ from the pre-main sequence up to the AGB phase. 
Contrary to previous calculations we found that these stars develop normal thermal pulses 
and third dredge-up episodes. Special attention is devoted to analyze the chemical 
enrichment in the envelope due to the above mechanisms. As a consequence, we show that these 
stars become carbon and nitrogen rich. U\-sing different IMFs proposed in the literature 
for the Population III stars, we study their contribution to the pre-galactic 
chemical enrichment. It is found that $Z=0$ AGB stars could significantly contribute 
to $^7$Li, $^{12}$C and $^{14}$N and produce extreme non-solar $^{24}$Mg/$^{25}$Mg/$^{26}$Mg 
ratios. However, the net contribution is very sensitive to the IMF adopted and to 
the fraction of primordial matter which goes into stars.
\end{abstract}

\section{INTRODUCTION}
The standard homogeneous Big Bang nucleosynthesis predicts that the material emerged from this 
epoch was mainly made by $^1$H and $^4$He. The total mass fraction of heavier elements was lower 
than $10^{-10}$, so that the first generation of stars, the so called Population III, was built 
from matter essentially deprived of metals. Such a peculiarity significantly affected the star 
formation phase. Yoshii and Saio \cite{YS86} found that the peak of the resulting IMF for Pop III stars ranges between 
4 and 8 $M_{\odot}$. Recently, Nakamura and Umemura \cite{NU99} have obtained that the typical mass of Pop III stars 
should be $\sim 3$ M$_\odot$. Nevertheless, recent simu\-lations of collapse and fragmentation of 
primordial clouds (see e.g. \cite{La99}) obtain a Jean mass $\sim 10^2-10^3$ M$_\odot$, although it 
is still not clear whether these clumps will further fragment down to stellar mass values. The key 
role in this process is played by molecular hydrogen rather than dust or heavy molecules cooling.
Therefore, early star formation with a normal present-day IMF seems very unlikely. In this framework, 
intermediate and/or very massive stars were the dominant constituents of the first stellar population. 

The simplest explanation for the absence of "zero metal" stars is that they have not survived up to 
the present epoch, which is compatible with the idea that primordial stars were not of low mass. Thus, 
the very metal poor stars presently observed were formed from matter already enriched by the ashes 
of {\it massive} Z=0 stars. Consequently, an indirect way to search for Pop III stars is to look 
for their nucleosynthetic imprint in the extremely metal poor objects now observed. As it is well known, 
the most important contribution of intermediate mass stars to the chemical enrichment of galaxies 
comes from the nucleosynthesis occurring during the AGB phase. 

\section{MODELS AND EVOLUTIONARY PROPERTIES}
The evolutionary models of 4, 5, 6, 7 and 8 M$_\odot$ (Z=0, Y=0.23) stars have been computed by the 
latest version of the FRANEC code \cite{Ch98}. No mass loss has 
been assumed. Owing to the lack of metals the main differences in the evolution with respect to that 
of metallic AGB stars are: 1) H burning starts at high temperature through the pp chains. The convective 
cores are smaller but H-burning extends to the 80$\%$ of the stellar mass. The core contracts until 
the $3\alpha$ reaction starts; then the burning switches to the CNO, 2) during He-burning the star 
is at the blue side of the H-R diagram, therefore the 1$^{st}$ dredge-up does not occur, and 3)
during the AGB phase the inward penetration of the envelope dredges-up He, C, N and O. For 
M$\geq 6$ M$_\odot$ the amount of CNO elements is enough to develop a normal TP phase. For 
M$<6$ M$_\odot$ the normal TP phase starts after an episode of C ingestion during the first pulses.
See \cite{St00} for more details.

\begin{table}[htb]
\caption{Important yields from Z=0 AGB stars.}
\label{table:1}
\newcommand{\m}{\hphantom{$-$}}
\newcommand{\cc}[1]{\multicolumn{1}{c}{#1}}
\renewcommand{\tabcolsep}{2pc} 
\renewcommand{\arraystretch} {1.2}
\begin{tabular}{@{}llllll}
\hline
IMF     &  \cc{$\eta$}   &\cc{$^4$He}  & \cc{$^7$Li} & \cc{$^{12}$C} & \cc{$^{14}$N}  \\
\hline
        
        &      1         &    0.1      & $2\times 10^{-10}$ & $2\times 10^{-3}$ & $4\times 10^{-3}$ \\
(a)     &                &             &                    &                   &                \\
        &      6         &    0.1      & $4\times 10^{-10}$ & $9\times 10^{-4}$ & $3\times 10^{-4}$       \\
        &      1         &    0.14     & $3\times 10^{-10}$ & $4\times 10^{-3}$ & $6\times 10^{-3}$        \\
(b)     &                &             &                    &                   &                     \\
        &      6         &    0.14     & $6\times 10^{-10}$ & $1\times 10^{-3}$ & $3\times 10^{-4}$   \\
        &                &             &                    &                   &                        \\
        &      1         &    0.02     & $4\times 10^{-11}$ & $4\times 10^{-4}$ & $7\times 10^{-4}$     \\
Salpeter&                &             &                    &                   &                   \\
        &      6         &     0.02    & $7\times 10^{-11}$ & $2\times 10^{-4}$ & $3\times 10^{-5}$    \\
\hline
\end{tabular}\\[2pt]
\end{table}

\begin{table}[htb]
\caption{Abundance ratios from Z=0 objects using the IMF by Yoshii \& Saio (1986).}
\label{table:2}
\newcommand{\m}{\hphantom{$-$}}
\newcommand{\cc}[1]{\multicolumn{1}{c}{#1}}
\renewcommand{\tabcolsep}{2pc} 
\renewcommand{\arraystretch} {1.2}
\begin{tabular}{@{}lllllll}
\hline
IMF   &           &   \cc{AGB}   &          &        & \cc{SNII}  &       \\
\hline
      & [C/Fe]    &              & [N/Fe]   & [C/Fe] &            & [N/Fe] \\
(a)   & +0.2      &              & +0.8     &  +0.2  &            & -0.45  \\
(b)   & +0.6      &              & +1.4     &  0.0   &            & -0.6   \\
\hline
\end{tabular}\\
\end{table}

\section{NUCLEOSYNTHESIS}
As a consequence of the occurrence of normal TP, $3^{th}$ dredge-up and 
 hot bottom burning, the Pop III nucleosynthesis 
scenario is revised. The duration of the whole AGB phase has been determined by using the
classical Reimers formula for the mass-loss rate for different values of the scaling factor
$\eta$. The major products of these stars are $^{4}$He, $^{7}$Li, $^{12}$C and $^{14}$N. In Table 1 
we report the most important (IMF weighted) yields provided by our Z=0 IMS models, namely: the abundance 
in mass fraction of a given element in the material ejected into the interstellar medium at the end 
of the AGB phase. We have used the Population III IMF proposed by Yoshii and 
 Saio \cite{YS86} (cases a \& b) and for 
comparison we also compute the yields from the Salpeter IMF. The [C,N/Fe] ratios are
obtained from Z=0 AGB and SN II yields assuming an ejection of 0.05 M$_\odot$ of Fe per SNII (Table 2). 
The main results are:

\begin{itemize}
\item Li is produced by the Cameron-Fowler mechanism particularly efficient during the
first part of the AGB phase. The maximum Li yield obtained is of the order of the standard
Big-Bang nucleosynthesis predictions.

\item C and N (primary) are built up during the AGB phase. However, the C and N yields
significantly change along the AGB, thus the cumulative amount of these elements finally  
ejected depend on the mass loss history. 

\item Free neutrons are released from the operation of the $^{22}$Ne$(\alpha,n)^{25}$Mg
reaction. The envelope is built up with extreme non-solar ratios $^{24}$Mg/$^{25}$Mg/$^{26}$Mg$=1/10/13$. 

\end{itemize}


\begin{thebibliography}{15}
\bibitem{Ch98} A. Chieffi, M. Limongi and O. Straniero 1998, ApJ, 737, 762
\bibitem{La99} R.B. Larson 1999, ESAB Symposium {\it Star Formation from
Small to Large Scale.} Eds. F. Favata, A.A. Kaas \& A. Wilson, p. 445
\bibitem{NU99} F. Nakamura and M. Umemura 1999, ApJ, 515, 239
\bibitem{St00} O. Straniero, I. Dom\'\i guez, A. Chieffi and M. Limongi 2000, ApJ (submitted)
\bibitem{YS86} Y. Yoshii and H. Saio 1986, ApJ, 301, 587 
 
\end{thebibliography}
\end{document}